\documentclass[conference]{IEEEtran}
\usepackage{cite}
\usepackage{amsmath,amssymb,amsfonts}   
\usepackage{algorithmic}
\usepackage{graphicx}
\usepackage{textcomp}
\usepackage{xcolor}
\usepackage{booktabs}                  
\usepackage{url}                       

\begin{document}

\title{Run, Read Back, Refine: Observation Fidelity in Execution-Grounded EDA Agents}

%
%
%
%
\newif\ifcamera
\cameratrue

\ifcamera
\author{%
\IEEEauthorblockN{Yang Liu, Shiwei Hou, Xiyuan Chen, Sen Yuan, Shao You, Feifan Chen, Wencheng Li, Shuyang Hu,\\
Emma Xia, Xiaojing Lu, Wenchuan Zhang, Hao Wang, Fan Xu, Yanfeng Li, Hongwen Li, Yiming Zhu}\\
\IEEEauthorblockA{\textit{Changxin Memory Technologies, Inc.}, China;\quad Email: \{paul.liu, yanfeng.li, howie.li\}@cxmt.com}
}
\else
\author{%
\IEEEauthorblockN{Anonymous Author(s)}
\IEEEauthorblockA{Paper under double-blind review}
}
\fi

\maketitle

\begin{abstract}
    Design engineers lose the most time orchestrating flows and patching the gaps
    where closed EDA tools refuse to interoperate. We present an
    execution-grounded LLM agent for PyAether, SKILL, and Tcl that closes a
    generate--execute--refine loop over unified MCP tools (API retrieval,
    documentation, sandbox execution) and is already deployed in production EDA
    flows, with a \emph{grounding-gap framework} positing that EDA-scripting
    correctness is bottlenecked by \emph{observation fidelity} at the L4
    (situated) layer, which model scale alone cannot reach. On
    \textbf{EDA-Scripting-Bench} (449 tasks, three languages, four
    commercial runtimes), the system reaches \textbf{76.5\%} pooled Pass@1 versus
    \textbf{19.7\%} (pure LLM): the system-level observation axis spans
    56.8 percentage points with the model fixed, exceeding the 23.3-percentage-point backbone range with
    the harness fixed---two axes answering different questions; observation is an under-invested axis whose returns are measurable and graded.
\end{abstract}

\begin{IEEEkeywords}
benchmarking, code generation, EDA scripting, execution grounding, LLM agents for EDA
\end{IEEEkeywords}

\section{Introduction}
\label{sec:introduction}

Design engineers' daily work is less about fresh scripts than about orchestrating design flows and patching the gaps between EDA tools that refuse to interoperate---a flow that cannot run end-to-end is the single largest source of lost design time. Commercial EDA tools are closed: where they leave a gap, the only recourse is hand-written glue (a Makefile target, a Tcl driver, a SKILL wrapper), precisely the small intent-to-code synthesis at which LLMs excel. This reframes EDA scripting as the connective tissue letting a general coding agent orchestrate flows and fill the breaks.

We present an execution-grounded LLM coding agent for PyAether, SKILL, and Tcl: four unified Model Context Protocol (MCP) tools close a generate--execute--refine loop over four commercial runtimes, in the spirit of recent MCP-based EDA agents \cite{lee2025autoeda}, with offline \textbf{API self-exploration} as a bounded complement. We also give a \textbf{theoretical grounding} (Section~\ref{sec:grounding-theory}): EDA correctness is situated (L4), a layer that offline knowledge alone cannot guarantee; the grounding gap on L4 failures is bounded by observation fidelity.



\noindent\textbf{Contributions.}
\begin{itemize}
    \item \textbf{RQ1 (binding constraint).} We formalize the grounding gap and measure both
    axes it depends on: with the LLM fixed, the system-level observation axis (pure LLM, no
    observation $\to$ full system) spans 56.8 percentage points (pp); with the harness fixed, the backbone axis
    spans 23.3~pp, while model scale still sets the L1--L3 ceiling
    (Section~\ref{sec:grounding-theory}, \ref{sec:experiments}).

    \item \textbf{RQ2 (necessary components).} Ablation shows retrieval and execution are statistically necessary (both effects exclude zero) and belief update is monotone in the execution budget, while self-exploration is a cost-negative prior-side aid rather than a substitute (Section~\ref{sec:experiments}).

    \item \textbf{RQ3 (toolchain generality).} \textbf{EDA-Scripting-Bench} (449 tasks, four
    runtimes, one schema) re-measures the same structure slice by slice; we specify the
    harness, schema, and assertion protocol in full (Section~\ref{sec:benchmark}).
\end{itemize}
\section{Background and Motivation}
\label{sec:background}

We target three widely-used EDA scripting interfaces across four commercial runtimes: PyAether for Empyrean Aether \cite{empyrean_pyaether}, SKILL for Cadence Virtuoso \cite{cadence2024virtuoso}, and Tcl for Synopsys Custom WaveView and Cadence Innovus. Despite their prevalence, scripting for these environments poses challenges beyond what generic LLM code generators can address.

\textbf{EDA scripting is fundamentally different.} Prior LLM code generation succeeds where correctness is verified by unit tests \cite{chen2021evaluating, austin2021program}. EDA scripting presents three distinct challenges: (i) APIs are long-tailed, tool-specific, and poorly documented; (ii) correctness depends on side effects in a live, stateful database, not return values; and (iii) the community lacks a standardized benchmark. Challenge~(ii) resists purely offline treatment: the live design state differs from what offline knowledge assumes.

\textbf{Prior custom/analog scripting work stops short of execution.} LayoutCopilot \cite{liu2025layoutcopilot} targets analog \emph{layout} (not scripting); ChaTCL \cite{rui2025chatcl} and RAG-EDA \cite{pu2024rag} generate Tcl scripts from static knowledge bases without executing them. Execution feedback exists in the digital flow (ChatEDA \cite{he2024chateda}; MCP4EDA \cite{wang2025mcp4eda} on Yosys/OpenLane; AutoEDA \cite{lee2025autoeda} on Tcl RTL-to-GDSII flows), but none grounds against the live layout, schematic, or waveform database of custom/analog tools. A recent survey \cite{liu2026agentic} organizes EDA agents by handoff validity; our system targets the custom/analog scripting handoff, previously ungrounded, and our benchmark is the first for this setting (concurrent benchmarks focus on digital Tcl \cite{xu2026iscript} or physical-design agents \cite{li2026pdagent}).

\textbf{Our approach.} We close the loop from generation to execution to refinement, supply a \textbf{theoretical grounding} (Section~\ref{sec:grounding-theory}) explaining \emph{why} execution is necessary, and provide the missing infrastructure: \textbf{EDA-Scripting-Bench} (Section~\ref{sec:benchmark}).
\section{System Design}
\label{sec:system}

\subsection{Overall Architecture}
\label{sec:overall-architecture}

\begin{figure*}[!t]
\centering
\includegraphics[width=\textwidth]{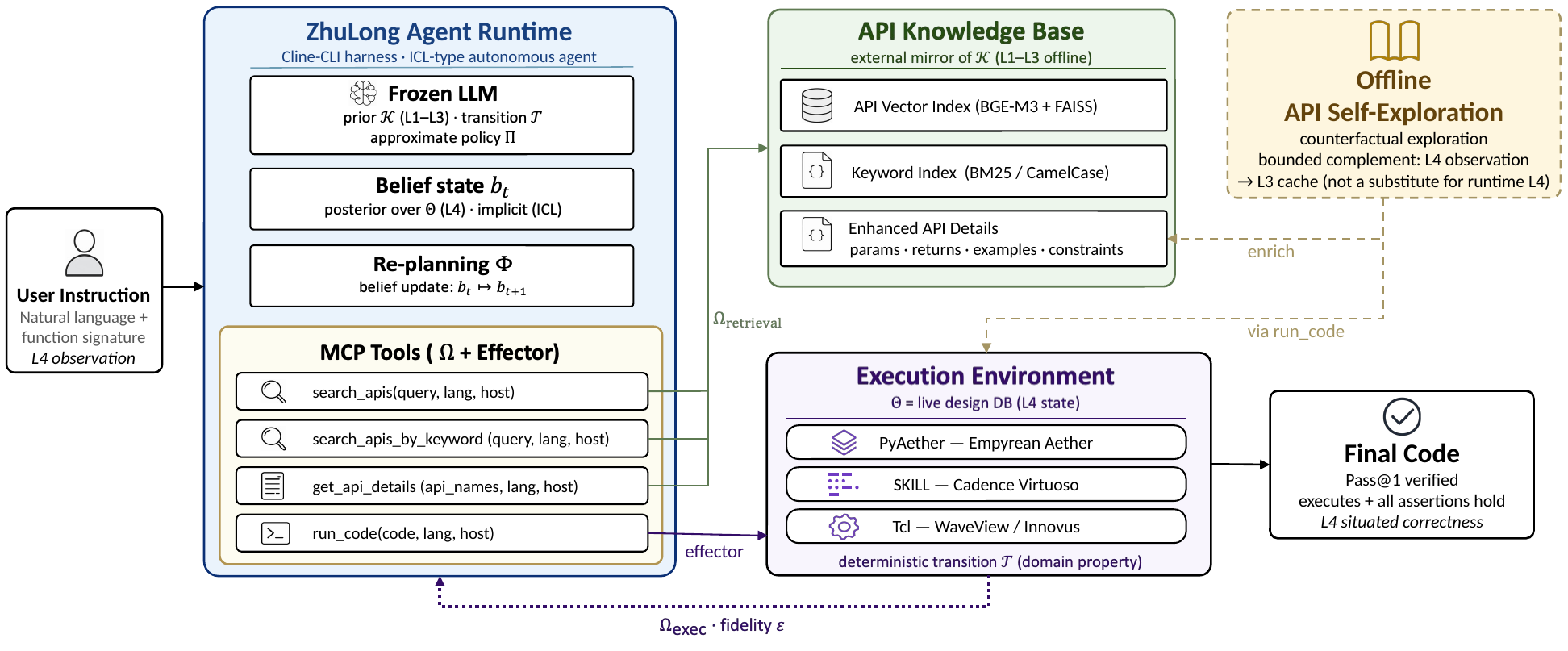}
\caption{System architecture. The LLM agent runtime, the API knowledge base, and the EDA execution sandbox are linked through \emph{four} unified MCP tools that dispatch on a single \texttt{lang}--\texttt{host} pair to \emph{three} platforms (PyAether, SKILL, Tcl). Offline API self-exploration pre-augments the knowledge base (dashed); execution feedback is looped back for iterative refinement.}

\label{fig:architecture}
\end{figure*}

Our system extends Cline-CLI \cite{cline2024cline} with EDA-specific retrieval, documentation, and execution. The runtime is an \textbf{ICL-type autonomous agent}: it decides its own next action, keeps no explicit posterior, and maintains its belief state implicitly via in-context learning (Section~\ref{sec:grounding-theory}). The policy stays commodity and unmodified, so gains are attributable to the interface, and the runtime's permission/hook mechanisms supply the enforcement points the fixed protocol relies on (Section~\ref{sec:exp-setup}). As in Fig.~\ref{fig:architecture}, three components---the LLM agent runtime, the API knowledge base, and the EDA execution environment---are linked through four MCP tools, with offline \textbf{self-exploration} pre-augmenting the knowledge base.

\subsection{Four MCP Tools}
\label{sec:mcp-tools}

The four tools jointly realize the observation channel $\Omega$ and the effector of Section~\ref{sec:grounding-theory}, each taking a \texttt{lang}--\texttt{host} pair that selects the runtime so one agent logic covers all four: \texttt{search\_apis} retrieves APIs by semantic similarity (BGE-M3 embeddings \cite{bge2024} with FAISS \cite{douze2024faiss}, top-$K$, merged with BM25); \texttt{search\_apis\_by\_keyword} retrieves by keyword (CamelCase plus BM25); \texttt{get\_api\_details} fetches documentation (parameters, returns, examples), augmented by self-exploration; and \texttt{run\_code} executes code in an isolated headless sandbox and reads back a structured result (status, error log, stdout, return) for refinement.

\subsection{Agent Workflow}
\label{sec:agent-workflow}

The agent iteratively refines its output through execution feedback \cite{cline2024cline}: reasoning over the instruction, retrieving and inspecting candidate APIs, generating and executing code, then re-planning based on the observed result. The agent is not constrained to a fixed sequence---it may skip retrieval when it already knows the API or conduct multiple rounds when initial results lack confidence---and terminates when it judges further attempts unlikely to succeed.

\subsection{Offline API Self-Exploration (Complementary)}
\label{sec:selfdoc}

EDA API documentation is often incomplete, omitting parameter constraints, return structures, and error conditions. Per Section~\ref{sec:grounding-theory}, retrieval enriches only the L1--L3 facts $\Omega$ returns, so self-exploration complements---and is bounded by---the L4 observation; it is an efficiency mechanism, not the L4 accuracy driver. Offline, the exploration agent reads documentation, identifies gaps, generates test scripts, executes them via \texttt{run\_code}, and iteratively refines a behavioral model of constraints, return structures, error conditions, and side effects. Exploration stops when a per-API confidence score exceeds threshold $\tau$, a round adds no new constraint, or a per-API budget $M$ is exhausted; scripts are expert-gated to filter destructive operations, and the result is served at runtime via \texttt{get\_api\_details} with no overhead. At collection time the exploration agent observes the live runtime via \texttt{run\_code}---an L4 observation---but the product is frozen into the knowledge base and served as a static fact, by then an L3-style record of a past observation, subject to staleness relative to the runtime state. This is why self-exploration is a bounded prior-side complement: it converts L4 observations into an L3 cache, and its accuracy is bounded by cache freshness, not by the fidelity of $\Omega$ at runtime.
\section{Theoretical Grounding: The Grounding Gap}
\label{sec:grounding-theory}

This section states the \emph{grounding gap} framework before Section~\ref{sec:experiments}'s measurements, as refutable predictions rather than fitted explanations: EDA correctness is a \emph{situated} property offline knowledge cannot guarantee, execution is the only channel observing the current design state, and on the L4 failures beyond offline reach the gap is bounded by observation fidelity---while model scale still sets the L1--L3 ceiling.

\subsection{A Four-Level Hierarchy of Information}
\label{sec:four-levels}

To separate what can be learned offline from what must be observed online, Table~\ref{tab:four-levels} distinguishes four levels: L1 \emph{syntactic correctness}, L2 \emph{physical plausibility}, and L3 \emph{historical factuality} (documented, past-oriented, frozen at the training cutoff) are learnable offline, whereas L4 \emph{situated truthfulness}---what is true \emph{right now} in a specific environment---is not, defined by indexicality (truth depends on context and changes over time), non-guaranteeability from offline data, and real-time commitment. L4 must be \emph{read} from the environment, not \emph{recalled} from memory.

\begin{table*}[!t]
\centering
\caption{The four levels of information}
\label{tab:four-levels}
\begin{tabular}{clp{5.2cm}c}
\toprule
\textbf{Level} & \textbf{Name} & \textbf{Core Question} & \textbf{Learnable Offline?} \\
\midrule
L1 & Syntactic Correctness & Is this string well-formed? & Yes \\
L2 & Physical Plausibility & Could this event happen? & Yes \\
L3 & Historical Factuality & Did this event happen? & Yes \\
L4 & Situated Truthfulness & Is this true \emph{right now}? & \textbf{No} \\
\bottomrule
\end{tabular}
\end{table*}

For EDA scripting, the situated state is the live design database---which cell/view is selected and what the layout or schematic contains now---plus the user's current intent (a user clarification is also an L4 observation). Our correctness criterion---code runs without error and all assertions over the mutated design pass---is L4: a well-formed (L1), plausible (L2) script invoking documented functions (L3) can still violate an undocumented binding contract, depend on internal selection state, or target the wrong cell. Such failures cannot be resolved offline; they must be \emph{observed} by acting on the live design and reading back the result. This paper's $\Omega$---retrieval plus execution read-back---selects one channel of L4, not its entirety: organizational, tacit, and modal situated information (waveform imagery, team conventions, un-digitized training knowledge) also falls under L4 but outside this $\Omega$.

Intuitively, L4 is what one can only know by being there: execution is the means of being there, but being there is not the same as seeing clearly---how clearly $\Omega$ sees is exactly $\epsilon$.\footnote{The L3/L4 boundary is a task-relative time window rather than an absolute clock: L3 is frozen at the training cutoff, while L4 is whatever falls inside the task's window and must be read from the environment.}

\subsection{A Bayesian Formalization of the Grounding Gap}
\label{sec:bayesian}

We cast grounding as decision-making under partial observability \cite{kaelbling1998planning, thrun2005probabilistic}; the formulation is diagnostic rather than algorithmic---it makes precise what is missing (the observation likelihood $\Omega$ and the recursive update $\Phi$) without prescribing tractable inference.

Let $\Theta$ denote the space of L4 states (the live EDA design database), $o_t$ a runtime observation, $a_t$ a generated code action, and $\mathcal{K}$ the offline knowledge (L1--L3) in the LLM's frozen parameters. The agent maintains a belief $b_t(\theta)\triangleq p(\theta_t\mid o_{1:t},\,a_{1:t-1},\,\mathcal{K})$ over the current L4 state, updated recursively by Bayes' rule:
\begin{equation}
b_t(\theta)\propto p(o_t\mid\theta_t)\!\int_{\Theta}\!p(\theta_t\mid\theta_{t-1},a_{t-1})\,b_{t-1}(\theta_{t-1})\,d\theta_{t-1},
\label{eq:pomdp}
\end{equation}
where $p(o_t \mid \theta_t)$ is the \emph{observation likelihood} $\Omega$ and $p(\theta_t \mid \theta_{t-1}, a_{t-1})$ is the \emph{transition model} $\mathcal{T}$ supplied by the LLM's prior; actions are chosen by an approximate policy $\Pi$ that minimizes expected cost under $b_t$, with $\mathcal{C}$ encoding the task assertions. \eqref{eq:pomdp} is the standard partially observable Markov decision process (POMDP) filter \cite{kaelbling1998planning}; the novelty is the \emph{implementation} of $\Omega$ as retrieval plus execution read-back.

The \emph{grounding gap} is the divergence between the posterior achievable with runtime observations and the posterior from the offline prior alone:
\begin{equation}
D_{\text{grounding}}(t) \triangleq D_{\mathrm{KL}}\!\left(p(\cdot \mid o_{1:t},\, \mathcal{K},\, \text{env}) \;\middle\|\; p_{\mathcal{K}}(\cdot)\right),
\label{eq:gap}
\end{equation}
where the right-hand side is the \emph{pure prior} unconditioned on any observation. We now bound this gap.

\textbf{Scenario: the transition is deterministic.} In EDA, \texttt{run\_code} executes a script whose effect on the design database is deterministic: the same action in the same state yields the same result, so $\mathcal{T}$ is exact---a property of the domain, not a simplifying assumption. The only mismatch is in the observation likelihood $\Omega$: retrieval may return incomplete documentation, and execution read-back may truncate---precisely what $\epsilon$ captures. Model mismatch under estimated parameters is a recognized obstacle \cite{delage2010percentile}; our setting is milder, with only $\Omega$, not $\mathcal{T}$, possibly mismatched.

\textbf{Claim (observation-fidelity bound).} Let $\epsilon \triangleq \sup_{\theta}\, \| p(\cdot\mid\theta) - \hat{p}(\cdot\mid\theta) \|_{\mathrm{TV}}$, the total-variation distance between the true and the agent-used observation likelihoods. We posit that $D_{\text{grounding}} \le g(\epsilon)$, with $g$ monotone non-decreasing and $g(\epsilon)\to 0$ as $\epsilon\to 0$. The bound is claimed on task distributions where L4 failures dominate: there, the grounding gap is bounded by observation fidelity; on L1--L3 failures, the prior $\mathcal{K}$ remains the relevant lever. We state this as a testable \emph{prediction}; the next paragraph tests it from both sides.

\textbf{Experimental correspondence (developed in Section~\ref{sec:experiments}).} Here $\epsilon$ is exactly the fidelity of this paper's $\Omega$ implementation---retrieval quality plus execution read-back. The read-back ladder (N$'$ $\to$ B $\to$ F, Table~\ref{tab:omega}) varies $\epsilon$ with retrieval and the effector fixed; the backbone axis (Table~\ref{tab:llm-comparison}) varies $\mathcal{K}$ with $\epsilon$ fixed. The two axes answer different questions---observation's return at a fixed model versus the prior's return under full observation---so we report both rather than as competing explanations. The read-back ladder isolates $\epsilon$ and is monotone on every slice; the system axis spans 56.8~pp while the backbone axis spans 23.3~pp; observation is the under-invested axis whose returns are measurable and graded. The bound is a limit statement; in deployment, however, $\epsilon \ge \epsilon_0 > 0$: by Section~\ref{sec:four-levels}, $\Omega$ covers one channel of L4, not its entirety, so the grounding gap keeps a positive lower bound in practice, and the 56.8-pp system span is correspondingly a lower bound on what the full grounded system buys---measured through the $\Omega$ this paper implements.

\subsection{Design Principles}
\label{sec:prediction}

Every execution-grounded harness already has a generate--execute--refine loop; a loop is necessary but not sufficient. What the loop alone leaves missing, and what this paper supplies, is a theory of what it cannot see: the L4 space no offline prior reaches, and the observation fidelity $\epsilon$ bounding how much of it a given $\Omega$ recovers. The formalization yields three architectural principles.
\textbf{Decouple knowledge from belief:} adapt in the dynamic posterior $b_t$ the Harness maintains, not by updating the frozen prior $\mathcal{K}$.
\textbf{Optimize observation, and recognize the prior's bounded reach on L4:} on L4 failures the gap is bounded by $\Omega$'s fidelity ($g(\epsilon)$ on the distributions where L4 dominates), so observation investment pays directly there---full observation beats binary feedback on both accuracy and operating cost (Section~\ref{sec:exp-results})---while a richer prior raises the L1--L3 ceiling but does not by itself close the L4 gap.
\textbf{Design for continuous belief updating:} $\Phi$ runs online ($b_t\mapsto b_{t+1}$); in the ICL-type agent (Section~\ref{sec:system}) what is bounded is not a round count but how much evidence belief can accumulate---a graded execution budget or a lagging read-back.
\section{Benchmark Construction}
\label{sec:benchmark}

\subsection{Dataset: EDA-Scripting-Bench}

\textbf{EDA-Scripting-Bench} is the instrument for both of this paper's comparisons: within one toolchain, the effect of each grounding component, and across toolchains, the effect of language structure. It covers PyAether, SKILL, and Tcl on four commercial runtimes, \textbf{449 tasks} in four slices (Table~\ref{tab:benchmark-composition}), all specified by one schema---a target language and runtime, a natural-language prompt, a function signature, and language-appropriate assertions---so a single harness (Section~\ref{sec:agent-workflow}) scores all four platforms uniformly. Scoring is execution-based (a task passes only when the generated code acts correctly in the live runtime), and the cross-language slices are difficulty-controlled by construction (Section~\ref{sec:benchmark-slices}).

\begin{table}[!t]
\centering
\small
\caption{Composition of EDA-Scripting-Bench}
\label{tab:benchmark-composition}
\setlength{\tabcolsep}{3pt}
\begin{tabular}{@{}llr@{}}
\toprule
\multicolumn{3}{@{}l}{\emph{Slices and runtimes}}\\
\textbf{Slice} & \textbf{Runtime} & \textbf{Tasks} \\
\midrule
EDA-Eval-PyAether & Empyrean Aether & 158 \\
EDA-Eval-Skill & Cadence Virtuoso & 157 \\
EDA-Eval-WaveViewTcl & Synopsys Custom WaveView & 77 \\
EDA-Eval-InnovusTcl & Cadence Innovus & 57 \\
\midrule
\multicolumn{2}{l}{Total} & \textbf{449} \\
\bottomrule
\end{tabular}
\end{table}

\subsection{The PyAether Slice (EDA-Eval-PyAether)}

The reference slice, \textbf{EDA-Eval-PyAether} (158 tasks), anchors every task to a concrete PyAether API; EDA experts curated each assertion to be self-contained and executable in a clean sandbox without vendor-internal context. Tasks span layout, schematic, and design-management workflows, compose several distinct APIs, and sometimes depend on a technology library, so passing reflects command of the tool's behavior, not surface retrieval.

\subsection{SKILL and Tcl Slices}
\label{sec:benchmark-slices}

The remaining slices are built so cross-language scores are comparable. \textbf{EDA-Eval-Skill} (157 SKILL tasks on Cadence Virtuoso) is \emph{function-aligned} with PyAether: each task re-implements the same entry point and functional specification as its PyAether counterpart, holding difficulty fixed. The \textbf{Tcl} slices are graded by difficulty across two further toolchains: waveform measurement on WaveView (\textbf{EDA-Eval-WaveViewTcl}, 77 tasks) and floorplanning, placement, and routing on Innovus (\textbf{EDA-Eval-InnovusTcl}, 57 tasks), so the cross-language comparison (Section~\ref{sec:cross-language}) attributes score differences to language structure rather than task hardness.

\subsection{Evaluation Protocol}

A task passes if the generated code executes in its target runtime sandbox without errors and all assertions hold; the primary metric is Pass@1 \cite{chen2021evaluating}, reported as mean $\pm$ standard deviation (s.d.) over five independent runs (reshuffled task orderings, fresh agent contexts), per slice and pooled over all 449 tasks.
\section{Experiments}
\label{sec:experiments}

\begin{table*}[!t]
\centering
\caption{Main results on EDA-Scripting-Bench: Pass@1 (\%) over five runs (mean $\pm$ s.d.), pooled over all 449 tasks and per slice. The five right-most columns report per-task operating cost over all 449 tasks (five runs, mean $\pm$ s.d.). The lower block varies the read-back $\Omega$ alone (retrieval and effector fixed); (F) coincides with the Full row.}
\label{tab:main-ablation}
\label{tab:omega}
\label{tab:cost}
\scriptsize
\renewcommand{\arraystretch}{0.82}
\setlength{\tabcolsep}{2pt}
\resizebox{\textwidth}{!}{%
\begin{tabular}{@{}lccccccccc@{}}
\toprule
 & \multicolumn{4}{c}{\textbf{Pass@1 (\%)}} & \multicolumn{5}{c}{\textbf{Per-task operating cost (449 tasks)}} \\
\cmidrule(lr){2-5} \cmidrule(lr){6-10}
\textbf{Configuration} & Overall (449) & PyAether (158) & SKILL (157) & Tcl (134) & Time (s) & run\_code & Tool calls & In tok (k) & Out tok (k) \\
\midrule
Pure LLM & 19.7$\pm$1.4 & 14.3$\pm$2.1 & 33.2$\pm$2.3 & 10.1$\pm$1.1 & 183$\pm$9 & --- & 3.6$\pm$0.1 & 72$\pm$2 & 6.3$\pm$0.7 \\
RAG (retr.+self-expl., no sandbox) & 68.6$\pm$1.6 & 71.1$\pm$4.6 & 52.2$\pm$1.9 & 84.6$\pm$1.9 & 226$\pm$9 & --- & 16.9$\pm$0.7 & 272$\pm$8 & 8.6$\pm$0.3 \\
Full w/o Retrieval & 66.2$\pm$4.2 & 73.4$\pm$7.3 & 45.2$\pm$9.4 & 82.2$\pm$2.3 & 981$\pm$151 & 22.6$\pm$2.8 & 26.3$\pm$3.1 & 658$\pm$133 & 11.7$\pm$2.8 \\
Full w/o Self-Expl. & 72.9$\pm$2.2 & 77.0$\pm$4.1 & 57.1$\pm$2.6 & 86.7$\pm$3.0 & 725.9$\pm$102.9 & 12.5$\pm$1.3 & 29.4$\pm$2.7 & 680.8$\pm$110.0 & 16.2$\pm$2.4 \\
\textbf{Full} & \textbf{76.5$\pm$1.0} & \textbf{82.5$\pm$2.0} & \textbf{59.1$\pm$1.6} & \textbf{89.9$\pm$2.0} & 550.4$\pm$43.3 & 9.8$\pm$0.5 & 25.6$\pm$1.3 & 636.2$\pm$54.4 & 11.5$\pm$1.0 \\
\midrule
\multicolumn{10}{@{}l}{\emph{Read-back axis (retrieval + effector fixed, only what} run\_code \emph{returns varies)}}\\
(N$'$) No read-back (prior only) & 68.8$\pm$1.9 & 73.4$\pm$2.1 & 50.7$\pm$4.9 & 84.5$\pm$3.8 & 273.6$\pm$4.5 & 1.1$\pm$0.0 & 16.6$\pm$0.8 & 277.3$\pm$11.0 & 4.0$\pm$0.1 \\
(B) Binary pass/fail & 71.9$\pm$1.8 & 74.6$\pm$1.6 & 57.2$\pm$4.8 & 86.0$\pm$2.2 & 681.6$\pm$57.3 & 22.2$\pm$3.5 & 43.2$\pm$4.0 & 848.3$\pm$146.5 & 13.5$\pm$1.3 \\
(F) Full observation (= Full row) & 76.5$\pm$1.0 & 82.5$\pm$2.0 & 59.1$\pm$1.6 & 89.9$\pm$2.0 & 550.4$\pm$43.3 & 9.8$\pm$0.5 & 25.6$\pm$1.3 & 636.2$\pm$54.4 & 11.5$\pm$1.0 \\
\bottomrule
\end{tabular}}

\scriptsize{Bold marks the per-column best; --- = no sandbox execution. Tcl pools WaveView (77) and Innovus (57) over 134 tasks. Time = net-active solve time (excluding quota wait); run\_code = execution calls per task; tokens per task (thousands), averaged over 449 tasks (five runs). Per-host Tcl: WaveView 88.6\%, Innovus 91.6\%.}
\end{table*}

\subsection{Experimental Setup}
\label{sec:exp-setup}

\textbf{Research questions.} Section~\ref{sec:grounding-theory} yields three testable claims---\textbf{RQ1} which axis binds, observation fidelity $\Omega$ or the prior $\mathcal{K}$ (S1, S3), \textbf{RQ2} whether each component is necessary and substitutable (S1, S2), and \textbf{RQ3} whether the structure reproduces on another toolchain (S4).

\textbf{Benchmark and protocol.} We use \textbf{EDA-Scripting-Bench} (449 tasks, Section~\ref{sec:benchmark}) with \textbf{Pass@1}: one trace per task (the agent may iterate via $\Phi$ but is never restarted), a 2,500-second timeout, and a \texttt{PreToolUse} hook enforcing a tool whitelist, blocking shell/code-search tools, and confining file access to a fresh directory---identical in every configuration. Each configuration runs five times (reshuffled order, fresh contexts), reported as mean $\pm$ s.d. No third-party agent can reach these runtimes without the MCP bridge, so baselines are tool-registration ablations of the same harness---enabling single-variable attribution. Tasks originate from non-redistributable vendor documentation.

\textbf{Configurations.} We isolate components by tool registration (a disabled tool is unregistered and invisible; the policy is fixed), giving the five rows of Table~\ref{tab:main-ablation}: \emph{Pure LLM} (no MCP tools), \emph{RAG} (retrieval-augmented generation: retrieval + self-exploration, no sandbox), \emph{w/o Self-Expl.} (retrieval + sandbox), \emph{w/o Retrieval} (sandbox), and \emph{Full} (all on). Main experiments use DeepSeek-V4-Pro; S3 evaluates five further backbones under the identical harness.

\subsection{Results and Ablations}
\label{sec:exp-results}

Table~\ref{tab:main-ablation} gives the five configurations pooled and per slice; against the pure-LLM floor the full system gains $56.8$~pp. Wilson 95\% intervals on the pooled Pass@1 ($n{=}2{,}245$) place Full at $[74.70,78.21]$ and Pure-LLM at $[18.11,21.40]$, and give the component gaps, all four of which exclude~0: retrieval $+10.3$~pp $[7.7,12.9]$, execution $+7.9$~pp $[5.3,10.5]$, self-exploration $+3.6$~pp $[1.1,6.1]$, and the binary-over-none read-back rung $+3.1$~pp $[0.4,5.8]$; only the SKILL-slice self-exploration difference $+2.0$~pp $[-2.9,6.9]$ remains within noise.\footnote{All difference intervals treat the two arms as independent and pool task-runs as i.i.d.; a paired, task-clustered analysis would only tighten them.}

\subsubsection{S1: Observation fidelity (RQ1)}
\label{sec:omega-fidelity}

Principle 2 predicts that, with the model fixed, accuracy is bounded by the fidelity of $\Omega$ \emph{on failures that require observing the live design}. \textbf{Documentation axis.} Holding the loop, effector, and LLM fixed, removing retrieval costs 10.3~pp pooled but is highly toolchain-specific. On PyAether the read-back partially substitutes for it---sandbox introspection via \texttt{dir()}/\texttt{inspect}/\texttt{\_\_doc\_\_} rises from 49.0\% to 86.7\% of traces once retrieval is removed---and Tcl substitutes likewise through \texttt{info commands} and \texttt{get\_db} probing, recovering the API surface but not its deep semantics ($7.7$~pp). SKILL's sandbox has no such channel (\texttt{apropos}/\texttt{describe} fail), only \texttt{boundp}/\texttt{isCallable} existence checks, so retrieval there is irreplaceable ($13.9$~pp). Retrieval is thus one \emph{implementation} of $\Omega$, not the channel.

\textbf{Read-back axis.} The second axis is how much live state execution returns. Three closed-loop variants differ only here (Table~\ref{tab:omega}, lower block): (N$'$) \emph{no read-back}---\texttt{run\_code} executes but returns nothing; (B) \emph{binary feedback}---pass/fail only; (F) \emph{full observation}---stdout, traces, per-assertion outcomes, and queried state (the default). The predicted ordering N$'$ $<$ B $<$ F holds pooled and on every slice. The middle rung is the smallest step ($+3.1$~pp pooled; $+1.2$~pp on PyAether, within one combined s.d.): a bare pass/fail verdict adds little actionable information, whereas the F rung carries most of the ladder ($+4.6$~pp of the $7.7$~pp N$'$-to-F gap). Cost sharpens the reading (Table~\ref{tab:cost}): binary feedback is \emph{more} expensive than full observation on every metric---without actionable error information the agent probes blindly---while no read-back collapses cost with accuracy. What the channel returns, not how often the agent acts, drives both---the N$'$-to-F gap operationalizes the grounding gap.

\textbf{Self-exploration rung.} The top rung adds constraints discovered offline (Section~\ref{sec:selfdoc}); per the framework this enriches only L1--L3 facts, so it should move the result less than the channel's fidelity. Removing it (retrieval and effector kept) drops Pass@1 from 82.5\% to 77.0\% on PyAether (76.5\% to 72.9\% pooled; Table~\ref{tab:main-ablation})---a bounded prior-side complement, not a substitute---and it is cost-negative (cutting net-active solve time by 24.2\%, Table~\ref{tab:cost}). Because this rung is served through \texttt{get\_api\_details}, \emph{w/o Retrieval} also drops it; the $+10.3$~pp retrieval gap thus subsumes its $+3.6$~pp gain.

\subsubsection{S2: The closed loop (RQ2)}
\label{sec:closed-loop}

With retrieval and self-exploration ablated in S1, we turn to the two remaining components. Principle 3 requires actions to reach the design through a causal interface and $\Phi$ to run continuously. \textbf{Effector.} Removing the sandbox drops pooled Pass@1 from 76.5\% to 68.6\% ($-7.9$~pp): without an effector the agent becomes a prior-only predictor. \textbf{Belief update.} We bound how often the agent \emph{may} execute (a per-trace budget denied at the tool boundary) rather than forcing early submission, varying evidence along two axes: the budget $k$ and a \emph{lagged} channel returning the \emph{previous} execution's state. On PyAether, accuracy rises monotonically with the budget (63.8\% at $k{=}1$, 69.1\% at $k{=}3$, 77.7\% at $k{=}10$, unsaturated); the lagged channel lands between $k{=}3$ and $k{=}10$ (73.0\%)---stale evidence is worth markedly less than fresh---and tighter budgets leave traces unfinished.

\subsubsection{S3: The prior axis (the prior-side contrast of RQ1)}
\label{sec:k-axis}

\begin{table}[!t]
\centering
\caption{Impact of LLM backbone (full harness).}
\label{tab:llm-comparison}
\scriptsize
\renewcommand{\arraystretch}{0.76}
\setlength{\tabcolsep}{2.5pt}
\begin{tabular}{@{}lccccc@{}}
\toprule
 & \textbf{PyAether} & \textbf{SKILL} & \textbf{Tcl} & \textbf{Overall} & \textbf{$\Delta$} \\
\textbf{Backbone} & (158) & (157) & (134) & (449) & \\
\midrule
DeepSeek-V4-Pro & 82.5$\pm$2.0 & \textbf{59.1$\pm$1.6} & \textbf{89.9$\pm$2.0} & \textbf{76.5$\pm$1.0} & --- \\
GLM-5.3-Flash & \textbf{86.8$\pm$1.6} & 57.2$\pm$6.9 & 80.3$\pm$3.8 & 74.5$\pm$2.4 & $-2.0$ \\
GLM-5.2 & 83.3$\pm$3.1 & 54.4$\pm$4.5 & 82.7$\pm$1.6 & 73.0$\pm$2.1 & $-3.5$ \\
Kimi-K2.6 & 77.0$\pm$1.6 & 55.5$\pm$2.7 & 75.8$\pm$1.9 & 69.1$\pm$0.5 & $-7.4$ \\
DeepSeek-V4-Flash & 60.1$\pm$5.5 & 47.0$\pm$1.8 & 70.4$\pm$3.1 & 58.6$\pm$2.3 & $-17.9$ \\
Doubao-Seed-2.0-Pro & 63.8$\pm$1.4 & 42.5$\pm$3.4 & 53.3$\pm$3.4 & 53.2$\pm$1.9 & $-23.3$ \\
\bottomrule
\end{tabular}

\footnotesize{Pass@1 (\%), five runs (mean $\pm$ s.d.); Tcl pools WaveView and Innovus over 134 tasks; $\Delta$ vs.\ the primary backbone. Bold = best per column; span $=76.5-53.2=23.3$~pp.}
\end{table}

Principle 1 separates knowledge from belief: the frozen prior sets the L1--L3 ceiling reachable without observation. Table~\ref{tab:llm-comparison} changes only $\mathcal{K}$ with the harness fixed; the backbone remains decisive---the strongest-to-weakest spread is what the $\mathcal{K}$ axis can move, raising the L1--L3 ceiling without reaching L4.

\subsubsection{S4: The toolchain axis (RQ3)}
\label{sec:cross-language}

The four slices share one schema, assertion criterion, and harness, so S1--S2 re-run per slice with only the toolchain changed. The pure-LLM floor and the closed-loop advantage reproduce on every slice, and the read-back ordering N$'$ $<$ B $<$ F holds pooled and per slice. Component magnitudes are toolchain-specific, tracking each host's introspective surface (slice-level: execution $+11.4$/$+6.9$/$+5.3$~pp on PyAether/SKILL/Tcl; retrieval $+13.9$/$+9.1$/$+7.7$~pp on SKILL/PyAether/Tcl), but the structure---observation plus execution closing the loop---reproduces everywhere.

\subsection{Error Analysis}
\label{sec:error-analysis}

The system passes 82.5\% on PyAether, 59.1\% on SKILL, 88.6\% on WaveView, and 91.6\% on Innovus (Table~\ref{tab:main-ablation}). Splitting misses across the five runs into \emph{generation-stage} (nothing executable returned) and \emph{evaluation-stage} (a candidate returned but failing) failures gives 29/109, 82/239, 1/43, and 1/23 on PyAether, SKILL, WaveView, and Innovus.

Failures reduce to two missing ingredients, not raw model capability: domain knowledge the prior needs for a candidate (L3), and a live-design contract $\Omega$ cannot see (L4). Generation-stage failures are predominantly L3---on SKILL, $73$ of the $82$ are timeouts, an execution-budget effect (the 2,500-second wall clock, not the $\Phi$ budget) rather than a knowledge gap---whereas the four patterns below are predominantly L4, constraints documented nowhere yet decisive, visible only through $\Omega$.

Across all four slices, the attributable evaluation-stage failures fall into four patterns---all on solvable tasks the system misreads through an underspecified or invisible contract: \textbf{(i) Wrong internal API or contract}---a plausible-but-wrong long-tail API or field literal; \textbf{(ii) Unauthorized defensive guards}---a self-added check rejecting input the specification never excluded; \textbf{(iii) Underspecified boundary and representation semantics}---an ambiguity resolved on the wrong side, flipping verdicts across runs; and \textbf{(iv) Host language and context contracts}---a type-template, headless-context, or canonical-form violation that kills the candidate before judging.

PyAether-068 uses the plausible-but-wrong \texttt{aeAddHigh\discretionary{-}{}{}LightRect} vs.\ the oracle's \texttt{GuCartBox(emyBoxF(...\discretionary{}{}{}))}; WaveView-037 returns \texttt{N/A} on a single-rising-edge waveform and SKILL-037 rejects an empty hierarchy path; Innovus-057 over-filters with an extra \texttt{.use=="signal"} predicate, reporting 1,869 weight-2 nets against the oracle's 1,872; and SKILL-008 hardcodes the schematic view type when opening a layout, missed every run though a direct probe shows it passable.

The four patterns explain the evaluation-stage misses but not SKILL's deficit, which stems from the generation stage---neither an L3 nor an L4 gap but an execution budget spent on unguided search. Under full observation SKILL's \texttt{error\_log} is the least informative: only 0.2\% of non-empty logs carry a line number and 2.3\% span more than one line, versus 88.6\% and 88.6\% for PyAether (median $67$ vs.\ $262$ characters)---a single \texttt{*Error*} naming the function, no site, no call stack; file-and-line reporting needs a debug flag, off by default---and even when enabled, the frames stay only as stderr text, outside the agent's read-back. Without a location cue the agent brute-force rewrites instead of repairing: timed-out traces issue a median of $23$ \texttt{run\_code} calls versus $4$ on passing traces, and 51\% of those probes return the $150$-second sandbox timeout. The timeouts are not unsolvable: $61$ of the $73$ timed-out traces share their task with a passing run (only $12$ never passed). Read-back fidelity is thus not only a policy knob (S1) but a toolchain property: weakest on SKILL, exactly where the component gaps bite (retrieval $+13.9$~pp---S4's largest slice-level magnitude---plus execution $+6.9$~pp).
\section{Discussion}
\label{sec:discussion}

Section~\ref{sec:grounding-theory} elevates the error analysis of Section~\ref{sec:error-analysis} to a structural principle.

Three limitations. \textbf{(i)} Base-LLM performance varies widely---consistent with the framework (the $\mathcal{K}$ axis moves L1--L3)---and the benchmark size limits power to certify the smallest effects (only the SKILL-slice self-exploration difference, $+2.0$~pp $[-2.9,6.9]$, remains within noise), so the self-exploration gain and the binary-over-none read-back rung are secondary contributions rather than headline findings. \textbf{(ii)} Per-slice dispersions can be large, so slice-level comparisons need care---the SKILL-slice self-exploration benefit is small, consistent with SKILL's smaller API surface. \textbf{(iii)} Pass@1 certifies only the effects its assertions check, not full functional equivalence, and it does not certify that the produced design passes layout-versus-schematic (LVS) or design-rule-check (DRC) checks, a gap this benchmark does not close. Separately, the system is deployed in production EDA flows; deployment metrics are omitted here.
\section{Conclusion and Future Work}
\label{sec:conclusion}

This work reframes EDA scripting as the connective tissue that makes design flows runnable end-to-end: a general coding agent orchestrates flows and synthesizes the glue where closed tools leave gaps. The grounding-gap framework explains why: the gap on L4 failures is bounded by observation fidelity, while scale still sets the L1--L3 ceiling. The full system reaches \textbf{76.5\%} pooled Pass@1. Future work: domain-specializing the harness, long-term memory and recursive self-improvement, multimodal grounding, and long-chain flow-orchestration benchmarking.

\clearpage
\bibliographystyle{IEEEtran}
\bibliography{reference}

\end{document}